\documentclass[12pt, prd, showpacs]{revtex4}
\usepackage{amssymb}
\usepackage{amsmath}
\usepackage{color}

\begin{document}

\title{Flow and peculiar velocities for generic motion in spherically
symmetric black holes}
\author{A. \ V. Toporensky}
\affiliation{Sternberg Astronomical Institute, Lomonosov Moscow State University }
\affiliation{Kazan Federal University, Kremlevskaya 18, Kazan 420008, Russia}
\email{atopor@rambler.ru}
\author{O. B. Zaslavskii}
\affiliation{Department of Physics and Technology, Kharkov V.N. Karazin National
University, 4 Svoboda Square, Kharkov 61022, Ukraine}
\affiliation{Kazan Federal University, Kremlevskaya 18, Kazan 420008 Russia}
\email{zaslav@ukr.net}

\begin{abstract}
In this methodological paper we consider geodesic motion of particles in a spherically symmetric black hole space-times.
 We develop an approach based on splitting the velocity of a freely falling  particle to the flow velocity,
 which depends only on a metric, and deviation from it (a peculiar velocity).     It applies to a wide class of spherically symmetric metrics and is exploited under the horizon of the Schwarzschild black hole. The present work generalizes previous results obtained for pure radial motion. Now, the motion is, in general, nonradial, so that an observer can have a nonzero angular momentum. This approach enables us to give simple physical interpretation of redshifts (blueshifts) inside the horizon including the region near the singularity and agrees with the recent results obtained by direct calculations.

\end{abstract}

\maketitle

\section{Introduction}

The goal of the present paper is to describe the geodesic motion in the
interior of a spherically symmetric black hole from a viewpoint which is
rather rarely presented in scientific and methodological literature
concerning black holes, and has been mostly developed in cosmology. The
concept of "expansion of space" is well known in methodological literature
and is widely used in cosmological textbooks (see, for example, 
\cite{Rindler}). Though from a purely
scientific perspective it is usually considered as a redundant concept, and
there are no formal differences in interpretations of Hubble expansion as
"expansion of space" and the motion of galaxies through the space (see \cite%
{Davis} and references therein), visualization of the Hubble flow and
peculiar velocities with respect to it can be considered to be pedagogically
useful. However, it rises a question of what is specific in cosmology that
gives rise to this concept. A recent development gives rather an unexpected
answer to this question: such an interpretation is not specific to cosmology
and can be developed in other areas of General Relativity.

In \cite{ham} it was suggested to think of a black hole as a river of space
that flows through it. Later, this picture was expanded in \cite{river2}, 
\cite{jcap}. In particular, this allows splitting of a velocity of a test
particle to two natural parts. One of them represents motion with a flow
while the second one gives deviation from it due to peculiar movement. This
leads to a rather transparent kinematic picture that is useful in a number of
application. For example, it was applied in \cite{jcap} to the description
of the so-called Ba\~{n}ados-Silk-West (BSW) effect \cite{ban} and in \cite{time}
to the problem of maximizing the time inside a black hole horizon.

Kinematics of particles (including the inner region of a black hole) was
considered in \cite{jcap} for pure radial motion only and for the
Gullstrand-Penlev\'e frame. 
Also, radial motion of particles in this frame (but without using the notions of flow and peculiar velocities) was considered in 
\cite{mx} for the Schwarzschild, Bardeen and Reissner-Nordsrtom black holes. 
Our goal here is to generalize our approach developed in \cite{jcap}  in two aspects. First, we consider an arbitrary motion that implies nonzero
angular momentum. Second, we use  synchronous coordinate systems. 
This allows us to make a direct comparison of our definitions with those usual in cosmology,
as well as to consider another (different from the Gullstrand-Penlev\'e one)
famous frame existing only inside a black hole horizon \cite{Novikov}.
Relying on this approach, we establish some generic
properties of motion near the black hole singularity and give simple
physical explanation of recent results for redshift inside a black hole \cite%
{20}.

We use the system of units in which fundamental constants $G=c=1$.

\section{Static spherically symmetric metric and equations of motion}

Let us consider the class of spherically symmetric metrics%
\begin{equation}
ds^{2}=-fdt^{2}+\frac{dr^{2}}{f}+r^{2}d\omega ^{2},\text{ }d\omega ^{2}=\sin
^{2}\theta d\phi ^{2}+d\theta ^{2}\text{.}  \label{met}
\end{equation}%
The horizon radius is $r=r_{+}$, $f(r_{+})=0$.

Eq.(\ref{met}) implies that $g_{tt} g_{rr}=-1$.
This form includes the Schwarschild, Reissner-Nordstr\"om, de Sitter and anti-de Sitter metrics, etc.
In principle, we can write a more general form with independent $g_{tt}$ and $g_{rr}$ but this causes technical difficulties without qualitative changes, so in the present paper we restrict ourselves to Eq. (\ref{met}).

We are interested in the behavior of test particles. For the spherically
symmetric case, motion occurs in a plane. We choose it to be $\theta =%
\frac{\pi }{2}$. Then, in the original frame (\ref{met}) the equations of motion
for a free particle with the energy $E$ and angular momentum  $L$ read:

\begin{equation}
\frac{dt}{d\tau }=\frac{\varepsilon }{f},  \label{te}
\end{equation}

\begin{equation}
\frac{dr}{d\tau }=-P,  \label{P}
\end{equation}%
\begin{equation}
\frac{d\phi }{d\tau }=\frac{\mathcal{L}}{r^{2}}\text{,}  \label{phi}
\end{equation}%
where $\varepsilon=\frac{E}{m}$, $\mathcal{L=}\frac{L}{m}$, $\tau$ is the proper time, $m$ is a particle mass and
\begin{equation}
P\equiv \sqrt{\varepsilon ^{2}-f(1+\frac{\mathcal{L}^{2}}{r^{2}})}\text{.}
\label{Z}
\end{equation}

We assumed that  a particle moves towards the horizon, so $r$ decreases with
time.

The original frame is deficient on the horizon where $f=0$. To make the
metric regular near the horizon, we apply the transformation%
\begin{equation}
t=\tilde{t}+F(r)-r^{\ast }\text{,}  \label{t}
\end{equation}%
\begin{equation}
r=\tilde{r}\text{,}
\end{equation}%
where%
\begin{equation}
r^{\ast }=\int^{r}\frac{d\bar{r}}{f(\bar{r})}
\end{equation}%
and $F$ is regular near the horizon $r=r_{+}$. Then, the metric can be rewritten in the new
coordinates in the form%
\begin{equation}
ds^{2}=-fd\tilde{t}^{2}+2d\tilde{t}dr(1-fF^{\prime })+dr^{2}[2F^{\prime
}-f\left( F^{\prime }\right) ^{2}]+r^{2}d\omega ^{2}.  \label{mtil}
\end{equation}%

Introducing also $v$ according to%
\begin{equation}
v=\sqrt{1-f}\text{,}  \label{vf}
\end{equation}%
and setting%
\begin{equation}
F^{\prime }=\frac{1-v}{f}=\frac{1}{1+v}\text{,}  \label{Ff}
\end{equation}%
we obtain the metric in the following famous form (the Gullstrand-Penleve (GP) metric)
\begin{equation}
ds^{2}=-d\tilde{t}^{2}+(dr+vd\tilde{t})^{2}+r^{2}d\omega ^{2}\text{.}
\label{GP}
\end{equation}

The transformation of time can be rewritten in the form
\begin{equation}
t=\tilde{t}-\int \frac{drv}{f}\text{.}
\end{equation}%
The form (\ref{GP}) is widely used in analog models of General Relativity, as well as in the so called
"river model" of black hole where $v$ represents a velocity of a flow ("river"), being a background on which 
standard Special Relativity considerations take place (see \cite{ham} for details).

For a massive particle, it follows from (\ref{te}) - (\ref{phi}) and (\ref{t}%
) that
\begin{equation}
\frac{d\tilde{t}}{d\tau }=\frac{\varepsilon -Pv}{f},  \label{tt}
\end{equation}%
\begin{equation}
\frac{dr}{d\tilde{t}}=-\frac{Pf}{(\varepsilon -Pv)},  \label{rti}
\end{equation}%
\begin{equation}
\frac{d\phi }{d\tilde{t}}=\frac{f}{\varepsilon -vP}\frac{\mathcal{L}%
}{r^{2}}.  \label{phit}
\end{equation}

Thus the four-velocity in coordinates $(\tilde{t},r,\phi )$ reads%
\begin{equation}
u^{\mu }=(\frac{\varepsilon -Pv}{f},-P,\frac{\mathcal{L}}{r^{2}})%
\text{.}  \label{ui}
\end{equation}

In the present paper we work also in synchronous frames. To obtain such a frame from the GP one,
 we transform not only the temporal coordinate but also a spatial one.
We want to obtain the metric in the form

\begin{equation}
ds^{2}=-d\tilde {t}^{2}+Ad\rho ^{2}+r^{2}(\rho ,\tilde{t})d\omega ^{2}
\end{equation}%
that would generalize the familiar Lema\^{\i}tre frame, well-known for the
Schwarzschild metric. 

Our goal is achieved with
\begin{equation}
\rho =t+\int^{r}\frac{d\bar{r}}{fv}.
\end{equation}%
Then,%
\begin{equation}
A=1-f=v^{2}\text{.}
\end{equation}%
If $f=1-\frac{r_{+}}{r}$, we return to the known formula
for the Lema\^itre form of the Schwarzschild metric
(see, for example, \cite{LL}).

The transformation presented here, from the explicitly
static form of the metric (1) to the Lema\^{i}tre
one (18), is a special case of a more general procedure
described in \cite{BR} and Sec. 3.3.3 of \cite{BRR}.

For free motion, using (\ref{tt}) and (\ref{P}) we have%
\begin{equation}
\frac{d\rho }{d\tau }=\frac{\varepsilon v-P}{vf}, \label{rotau}
\end{equation}%
\begin{equation}
\frac{d\rho }{d\tilde{t}}=\frac{\varepsilon v-P}{(\varepsilon -Pv)v}.
\label{rotilda}
\end{equation}

\section{Definitions of peculiar velocity}

Now, we want to describe particle motion by different types of observers. To
this end, we attach a tetrad to an observer, where the role of a time-like
vector is played by a four-velocity of a reference observer. In particular,
if we choose such an observer as a free falling one, with special values of
the energy and angular momentum, this defines a flow. Deviations from it
correspond to a peculiar motion. If we choose a suitable definition of a
three-velocity $V^{(i)}$ ($i=1,2,3)$, such a reference observer will have $%
V^{(i)}=0$ whereas a peculiar motion is described by nonzero $V^{(i)}$. In
other words, $V^{(i)}$ describes a motion of a particle with respect to the
flow.

To choose correct and physically reasonable definition of $V^{(i)}$, we use
the tetrad formalism. Then, the motion of a massive particle is subluminal, so
the absolute value of the vector $V^{(i)}$ is less than $1$. In this
section, we compare two different definitions of such a velocity which were
used independently in different contexts and show that they coincide.

Let \ a particle and an observer have four-velocities $u^{\mu }$ and $\tilde{u}%
^{\mu }$ correspondingly. Then, following eq. (2) of \cite{bol}, we can
introduce the quantity $w^{\mu }$ according to

\begin{equation}
w^{\mu }=\tilde{u}^{\mu }+\frac{u^{\mu }}{(u,\tilde{u})}\text{.} \label{w}
\end{equation}

Eq. (\ref%
{w}) was used, in particular, in eq. (40) of \cite{rad}. Note that our notations of four-velocities 
differ from those in \cite{bol, rad}.
The sign at the second term in (\ref{w}) differs from that in \cite{rad} because of our signature $%
(-,+,+,+)$ instead of $(+,-,-,-)$ in \cite{rad}. Here, $(u,u)=(\tilde{u},%
\tilde{u})=-1$ instead of $+1$ in \cite{rad}.

Meanwhile, there is a standard definition of three-velocity (see, eq. 3.9 in
e.g. \cite{72})%
\begin{equation}
V^{(i)}=-\frac{u^{\mu } h_{(i)\mu}}{h^{\mu (0)}u_{\mu }}.
\label{V}
\end{equation}

Let us choose the tetrad of basis vectors. We choose vector "0" along the
four-velocity,%
\begin{equation}
h_{(0)}^{\mu }=\tilde{u}^{\mu }\text{, }h^{\mu (0)}=-\tilde{u}^{\mu }. \label{h0}
\end{equation}%
Then, we see that%
\begin{equation}
w_{\mu }h_{(0)}^{\mu }=0\text{.}
\end{equation}%
Other vectors are orthogonal to "0" by construction, so%
\begin{equation}
h_{(i)\mu }h^{\mu (0)}=0\text{.}  \label{i0}
\end{equation}%
Then, it follows from (\ref{w}) and (\ref{i0}) that%
\begin{equation}
w_{\mu }h_{(i)}^{\mu }=\frac{u^{\mu }h_{(i)\mu }}{(u,\tilde{u})}
\end{equation}%
The denominator can be rewritten using (\ref{h0}),%
\begin{equation}
(u,\tilde{u})=h_{(0)}^{\mu }u_{\mu }=-h^{\mu (0)}u_{\mu }.
\end{equation}%
Then, we obtain that%
\begin{equation}
w_{\mu }h_{(i)}^{\mu }=V_{(i)}=V^{(i)}
\end{equation}%
coincides with (\ref{V}) as it should be.

We can define the tetrad components of $w^{\mu }$ according to 
\begin{equation}
w^{(a)}=h^{(a)\mu }w_{\mu }.
\end{equation}

Then, we have%
\begin{equation}
w^{(a)}=(0,V^{(i)})\text{.}
\end{equation}

\section{Local and non-local velocities in a synchronous frame}
The concept of a peculiar velocity is widely used in cosmology where it usually means a velocity with respect to the FLRW  frame. Since the flat FLRW metric is 
\begin{equation}
ds^2= -dt^2 + a(t)^2 d\chi^2 + a(t)^2 \chi^2 (d\theta^2+\sin^2{\theta} d\phi^2)
\end{equation}
($a$ is the scale factor and $\chi$ is   the radial comoving coordinate)
the corresponding tetrad is
\begin{equation}
h_{(i) \mu}=diag(-1, a,a \chi ,a \chi \sin{\theta})
\end{equation}
and for the 3-velocity of a particle with 4-velocity $u^{\mu}$ we have according to (\ref{V})
\begin{equation}
V_i=\frac{h_{(i) \mu} u^{\mu}}{-h_{(0) \mu} u^{\mu}}.
\end{equation}
In particular, the radial component of a peculiar velocity 
\begin{equation}
V_r= a u^{\chi}/ u^t = a d \chi/dt. 
\end{equation}

The consent of a peculiar velocity is a local one, however, the radial component allows for a nice
non-local interpretation \cite{Davis2}. Namely, the rate of change of a proper distance between the coordinate origin and a distant point
$l=a \chi$ is 
\begin{equation}
\frac{dl}{dt} = \frac{d(a \chi)}{dt} = \chi \frac{da}{dt} + a \frac{d \chi}{dt} = v_{H} + V_r
\label{sum}
\end{equation}
where $ v_{H} = \chi \dot a = (\chi a) \dot a/a = l H$ is the velocity of the Hubble flow. The left hand side
of this equation can be considered as a reasonable  definition of a velocity of a {\it distant} object (more precisely,
its radial part) which is an intrinsically non-local entity.
The above equation means that the overall change of a proper distance to a distant point is naturally decomposed into 
a sum of the velocity of the cosmological flow and the radial part of a peculiar velocity. Note that
summation rule is of the  Galileo type independently of the velocity values. We see that in spite of  a non-local nature of this equation
as a whole, as well as the Hubble flow velocity $v_H$, the second term in the right hand side has a local interpretation.

It is also worth noting  that the rate of change of a proper distance to a remote point is
a natural definition only for one (radial) component of the velocity of a distant point. Obviously, this rate is useless
for defining distant tangent velocities. There are several proposals to define them, which 
give different results, as well as other definitions of a radial velocity,
see, for example, \cite{bol, Chodorowski, Emtsova}. We will not consider tangent non-local
velocities and other definitions of a radial non-local velocity in the present paper.

In \cite{we} it was noticed that the property (\ref{sum}) is not specific to cosmology and is present in any
spherically symmetric synchronous system. Indeed, if the value of the “scale factor” $a$ depends on $\chi$ as well, and the particle has the comoving coordinate $\chi_1$,
we get the same result using the Leibniz integral rule:

\begin{equation}
\frac{dl}{dt}=\frac{d}{dt}\int_0^{\chi_1} a(\chi,t) d\chi= \int_0^{\chi_1} \frac{da(\chi,t)}{dt} d\chi+
a(\chi_1,t) \frac{d \chi}{dt} = v_{H} + V_r
\label{gensum}
\end{equation}

The radial velocity in the local sense is the same since the derivation of (36) does not 
use the spatial homogeneity. Moreover, both derivations (the local and non-local ones)
can be valid even if we abandon spherical symmetry, but are still able to introduce
a coordinate system in which the free flow is one-dimensional — in this case 
non-radial components of the metric (whatever they be)  do not enter in the derivation.

In the present paper we mainly describe properties of peculiar velocities in the Schwarzchild space-time.
This concept  allows us to get different interpretations of the known physical effects near and inside 
a black hole horizon. In our previous paper \cite{jcap} we already used the radial part of the peculiar velocity to reinterpret
BSW-like effects, giving a different perspective (particles which are "slow" in a conventional  sense appear to 
be "fast" in the sense of radial peculiar velocity and {\it vice versa}). Now we extend our analysis to angular peculiar
velocities and show how the known results on redshift observed inside a horizon can be explained in this approach.

\section{Static observer}
Although our ultimate goal in the present paper is to consider 3-velocities with respect to synchronous frames,
we start with a more common case of a static observer.

Then, in the original coordinates $(t,r, \theta,\phi)$%
\begin{equation}
e_{(0)\mu }=-(\sqrt{f},0,0,0),  \label{st0}
\end{equation}%
\begin{equation}
e_{(1)\mu }=(0,\frac{1}{\sqrt{f}},0,0).  \label{st1}
\end{equation}%
\begin{equation}
e_{\mu (2)}=r(0,0,1,0)\text{,}
\end{equation}%
\begin{equation}
e_{\mu (3)}=r\sin \theta (0,0,0,1)\text{,}
\end{equation}%

Then, the motion of an observer with respect to this frame within the plane $\theta=\pi/2$ is characterized
by the three-velocity with the tetrad components%
\begin{equation}
V_{st}^{(1)}=\frac{u^{r}}{fu^{t}}=\frac{1}{f}\frac{dr}{dt}=-\frac{P}{%
\varepsilon }  \label{vst1}
\end{equation}%
\begin{equation}
V_{st}^{(3)}=r\frac{u^{\phi }}{\sqrt{f}u^{t}}=\frac{r}{\sqrt{f}}\frac{d\phi 
}{dt}=\frac{\mathcal{L}\sqrt{f}}{r\varepsilon }.  \label{vst3}
\end{equation}%
Now, for $V_{st}^{2}=V_{st}^{(1)2}+V_{st}^{(3)2}$ we have 
\begin{equation}
V_{st}^{2}=\frac{P^{2}+\frac{\mathcal{L}^{2}}{r^{2}}f}{\varepsilon ^{2}}=%
\frac{\varepsilon ^{2}-f}{\varepsilon ^{2}}\text{.}  \label{vst}
\end{equation}

\section{Observer in the Lema\^itre frame}

Now, we shall consider free falling observers. Let $\varepsilon=1$ and $\mathcal{L}=0$. We will call this observer
the  Lema\^itre frame observer for brevity, though we mostly use the $(\tilde t, r)$ coordinates of the GP metric (\ref{GP}) in this section.
For such an observer, in the original coordinates 
$(t,r)$%
\begin{equation}
U^{t}=\dot{t}=\frac{1}{f}\text{, }
\end{equation}%
\begin{equation}
U^{r}=\dot{r}=-\sqrt{1-f} =-v\text{.}
\end{equation}%
 Here we used (1) and (3), (5) with $\varepsilon=1$, $L=0$.
We denote a corresponding tetrad $h_{(a)}^{\mu }$. In the coordinates $(t,r, \theta, \phi)$ it
has the form%
\begin{equation}
h_{(0)\mu }=(-1,-\frac{v}{f},0,0),
\end{equation}%
\begin{equation}
h_{(1)\mu }=(v\text{, }\frac{1}{f},0,0),
\end{equation}%
\begin{equation}
h_{(3)\mu }=r(0\text{, }0,0,1).  \label{h3}
\end{equation}%
In the coordinates $(\tilde{t},r, \theta, \phi)$, we have%
\begin{equation}
h_{(0)\mu }=(-1,0,0,0),  \label{h0m}
\end{equation}%
\begin{equation}
h_{(1)\mu }=(v\text{, }1,0,0).
\end{equation}%
Let some particle moves with the four-velocity $u^{\mu }$. By the definition (24),%
\begin{equation}
V^{(i)}=-\frac{h_{(i)\mu }u^{\mu }}{h_{(0)\mu }u^{\mu }}.  \label{vb}
\end{equation}

After a simple calculation we obtain that $V^{(1)}$ satisfies
\begin{equation}
\frac{dr}{d\tilde{t}}=-v+V^{(1)}\text{,}  \label{spl}
\end{equation}%
thus returing to eq. 5.3 of \cite{jcap}.
This gives us an intuitive interpretation of a full radial velocity (the rate of change of the radial
coordinate $r$ with respect to the Lema\^itre time $\tilde t$) as a sum of the flow velocity taken with negative
sign and a radial peculiar velocity. Note the difference between (\ref{spl}) and 
the general result of this type  for synchronous systems (\ref{gensum}), which is still valid if we change
the notation for the radial spatial variable from $\chi$, usually used in cosmology,  to $\rho$ as it is accepted
for the Lema\^itre metric (18).
Namely, the general formula (\ref{gensum}) with the same form of the  right hand side as in (\ref{spl})
deals with the rate of change of a proper distance $l$, not the radial coordinate $r$. However,
 in the Lema\^itre  frame the proper distance 
between two points at the same radius is equal simply to the difference of their values of $r$ -- we can see from (12) that
sections of constant time $\tilde t$ are flat.
That is why in this particular frame the radial coordinate $r$ (instead of such a nonlocal entity as the proper distance $l$) appears  in the left hand side of (\ref{spl}),  and 
the Lema\^itre frame appears to be especially useful in visualizing both inside and outside regions of a black hole in
a single picture.
The sign of $v$ here should be reversed (in contrast to the cosmological case) since the free flow motion is directed towards lower values of $r$, while in the cosmological case the Hubble flow makes the radial distances increase.

For a particle in the flow $V^{(1)}=0$.
In a general case, when peculiar motion is present, Eq. (\ref{rti}) gives us 
\begin{equation}
V^{(1)}=\frac{\varepsilon v-P}{\varepsilon -Pv}.  \label{v1p}
\end{equation}
Taking also into account (\ref{rotau}) and (\ref{rotilda}) one can check
that $v d\rho/d\tilde{t}$ (the radial peculiar velocity of (\ref{gensum})) coincides with (\ref{v1p}), as  should be the case.

The inverse formula reads%
\begin{equation}
P=\varepsilon \frac{v-V^{(1)}}{1-vV^{(1)}}\text{.}  \label{PVv}
\end{equation}
These formulae relate the radial peculiar velocity with the integrals of motion.

For the component $V^{(3)}$ we obtain from (\ref{h0m}), (\ref{h3}), (\ref{vb}%
) that%
\begin{equation}
V^{(3)}=r\frac{u^{\phi }}{u^{\tilde t}}=r\frac{d \phi}{d \tilde t}
\text{.}
\end{equation}%
The reason why this formula is so simple is again the fact that the sections of constant
time $\tilde t$ are flat, and we can use standard relations of Euclidean geometry.

For free motion, using (\ref{phit}) and (\ref{rti}), we have 
\begin{equation}
V^{(3)}=\frac{\mathcal{L}}{r}\frac{1-v^{2}}{\varepsilon -vP}=\frac{\mathcal{L%
}}{r}\frac{f}{\varepsilon -vP}.  \label{v3f}
\end{equation}

From (\ref{v1p}) and (\ref{v3f}) we can obtain the relation that will be
used below:%
\begin{equation}
V^{(3)2}=\frac{\frac{\mathcal{L}^{2}}{r^{2}}}{1+\frac{\mathcal{L}^{2}}{r^{2}}%
}(1-V^{(1)2})
\end{equation}%
whence%
\begin{equation}
\frac{1}{1-v_{p}^{2}}=\frac{1+\frac{\mathcal{L}^{2}}{r^{2}}}{1-V^{(1)2}}%
\text{,}  \label{vp}
\end{equation}%
where 
\begin{equation}
v_{p}^{2}\equiv V^{(1)2}+V^{(3)2}.  \label{pec}
\end{equation}

Now it is possible to express the energy $\varepsilon$ through the components of the peculiar 
velocity. Indeed,
it follows  from (\ref%
{PVv}) that
\begin{equation}
\varepsilon (v-V^{(1)})=P(1-vV^{(1)}).
\end{equation}

Taking the square and recalling (5), we obtain%
\begin{equation}
\varepsilon ^{2}=\frac{(1-vV^{(1)})^{2}}{1-v_{p}^{2}}\text{,}
\end{equation}%
where (\ref{vp}) was used.
Thus,
\begin{equation}
\varepsilon =\frac{1-vV^{(1)}}{\sqrt{1-v_{p}^{2}}}\text{.}  \label{evp}
\end{equation}
Here, we chose the sign to have $\varepsilon >0$ in the limit $v\rightarrow 0
$.

Note, that independently of the angular velocity, $\varepsilon =0$ always
corresponds to $vV^{(1)}=1$. We will consider such particles further in the
next section. They can exist only inside the horizon where $v$ exceeds the
speed of light, so $V^{(1)}$ is subliminal, as it should be for a velocity
having a direct physical meaning. Equation (\ref{evp}) gives us also a simple
criterion for the energy to be negative in terms of the flow and radial peculiar
velocity: $\varepsilon <0$ is equivalent to $V^{(1)}>1/v$.

Also, from (\ref{PVv}) and (\ref{evp}) a useful relation follows:%
\begin{equation}
P=\frac{v-V^{(1)}}{\sqrt{1-v_{p}^{2}}}\text{.}  \label{PV1}
\end{equation}

Then, one can also rewrite (\ref{v3f}) as%
\begin{equation}
V^{(3)}=\frac{\mathcal{L}}{r}\sqrt{1-v_{p}^{2}}.  \label{v3v}
\end{equation}

To finish this section, we also relate velocities with respect to the Lema\^itre and static frames.

As follows from (\ref{vst1}) - (\ref{vst}) and (\ref{v1p}),
\begin{equation}
V^{(1)}=\frac{v+V_{st}^{(1)}}{1+vV_{st}^{(1)}}. \label{sm}
\end{equation}%
In the case under discussion, $V_{st}^{(1)}$ is proportional to $-P$ according to (43), so $%
V_{st}^{(1)}$\thinspace $<0$ (the motion is towards a black hole). We can recognize in (\ref{sm}) the Lorentz composition
law for collinear velocities $v$ and $V_{st}^{(1)}$.

Also, taking into account (\ref{vst3}) we obtain%
\begin{equation}
V^{(3)}=V_{st}^{(3)}\frac{\sqrt{f}}{1+vV_{st}^{(1)}}=V_{st}^{(3)}\frac{\sqrt{1-v^2}}{1+vV_{st}^{(1)}},
\end{equation}
which represents the Lorentz composition law for perpendicular velocities.

The inverse formulas read%
\begin{equation}
V_{st}^{(1)}=\frac{V^{(1)}-v}{1-vV^{(1)}}\text{,}
\end{equation}%
\begin{equation}
V_{st}^{(3)}=\frac{V^{(3)}\sqrt{f}}{1-vV^{(1)}}.
\end{equation}

\section{Limiting transitions}

Let us consider the horizon limit, when $f\rightarrow 0$, $v\rightarrow 1$.
 We first consider the case of $\varepsilon>0$. Then,
\begin{equation}
P=\varepsilon -\frac{f}{2\varepsilon }(1+\frac{\mathcal{L}^{2}}{r_{+}^{2}}%
)+...,
\end{equation}%
\begin{equation}
v=1-\frac{f}{2}+...
\end{equation}%
Then, it follows from (\ref{v1p}) and (\ref{v3f}) that%
\begin{equation}
V^{(1)}\rightarrow \frac{1-\varepsilon ^{2}+\frac{\mathcal{L}^{2}}{r_{+}^{2}}%
}{1+\varepsilon ^{2}+\frac{\mathcal{L}^{2}}{r_{+}^{2}}} \label{limr}
\end{equation}%
\begin{equation}
V^{(3)}\approx \frac{2\mathcal{L\varepsilon }}{r_{+}(\varepsilon ^{2}+1+%
\frac{\mathcal{L}^2}{r_{+}^{2}})} \label{limphi}
\end{equation}

We remind a reader that the velocity with respect to a static frame at the
horizon always has components in this limit (see Eqs. (43), (44))%
\begin{equation}
|V_{st}^{(1)}|\rightarrow 1\text{,}  \label{stc}
\end{equation}%
\begin{equation}
V_{st}^{(3)}\rightarrow 0\text{.}
\end{equation}

On the contrary, the velocity with respect to the Lema\^{\i}tre frame at the
horizon does depend on the particular motion of the particle in question. In
particular, the radial velocity can take any value in the range 
$0 \le |V^{(1)}| <1$. For  pure radial motion  the left limit is realized  for $\varepsilon =1$, and this corresponds to
a particle comoving with the Lema\^{\i}tre frame, so the peculiar velocity
is evidently zero. The upper limit is a  limiting case for $%
\varepsilon \rightarrow 0$. As for the angular part of the peculiar velocity,
near the horizon $V_{st}^{(3)}$ is small, so a particle hits a horizon radially. However, a small $V_{st}^{(3)}$ near the horizon is compensated by a significant Lorentz boost (68) with (75). As a result, in the Lemaitre system $V^{(3)}$ is finite and nonzero. It is  a bright manifestaton of the known relativistic effect according to which a vector, not collinear to the direction of motion, rotates under the Lorentz transformation.

If the metric has an inner horizon, 
 the case of $\varepsilon<0$ for a particle approaching the horizon is  possible as well. The asymptotics
for a negative  energy are totally different, since now 
\begin{equation}
P \to -\varepsilon, \label{minus}
\end{equation}
and Eq.(16) gives us
$d\phi/d\tilde t \to 0$, so that 
\begin{equation}
    V^{(3)} \to 0.
\end{equation}
As for the radial motion, (\ref{v1p}) in the limit (\ref{minus}) gives 
\begin{equation}
    V^{(1)} \to 1,
\end{equation}
so that $v_p \to 1$.

Near the singularity, $r \to 0$,  $f\rightarrow -\infty $,
and for  pure radial motion%
\begin{equation}
    v\approx  P \approx \sqrt{\left\vert f\right\vert } 
\end{equation}
which gives us from (\ref{v1p}) that
\begin{equation}
V^{(1)}\approx \frac{1-\varepsilon}{v}\approx \frac{1-\varepsilon}{\sqrt{\left\vert f\right\vert }}%
\rightarrow 0\text{.}    
\end{equation}
For a non-radial motion with $\mathcal{L} \ne 0$, we have a different asymptopics of $P$
\begin{equation}
P\approx \sqrt{%
\left\vert f\right\vert }\frac{\mathcal{L}^{2}}{r^{2}}\gg v
\end{equation}%
and it follows from (\ref{v1p}) that
\begin{equation}
V^{(1)}\approx \frac{1}{v}\approx \frac{1}{\sqrt{\left\vert f\right\vert }}%
\rightarrow 0\text{.} \label{v1l}
\end{equation}%
This means that any initial differences in radial motion for different
particles disappear near the singularity, and the radial motion of any
particle tends to the motion of the frame. On the contrary, if $\mathcal{L}%
\neq 0$, then%
\begin{equation}
V^{(3)}\approx \frac{\mathcal{L}}{\left\vert \mathcal{L}\right\vert }=\pm 1%
\text{,} \label{v3l}
\end{equation}%
so pure radial motion appears to be unstable - an arbitrary small
deviation grows infinitely and results in ultrarelativistic motion in
angular direction. If initially the directions of the
vectors $\mathcal{L}$ are distributed randomly, the corresponding particles have
mutual ultrarelativistic relative velocities near a singularity.

\section{Another synchronous system inside a horizon}

In this section we consider another synchronous system (see, for example, \cite{Novikov} and page 25 
of the book \cite{Frolov}), which can be
obtained from static coordinates when using them inside the horizon. In this
region the initial signature of the metric changes, and the coordinate $r$
becomes timelike, as well as the coordinate $t$ changes its nature and
becomes spacelike. Therefore, it is appropriate to make a simple
change of notations defining a new time coordinate $T=-r$, a new spatial
coordinate $y=t$ and a function $g=-f$. Note, that $g$ is a function of the new
time inside the horizon.

\begin{equation}
ds^{2}=-\frac{dT^{2}}{g}+gdy^{2}+T^{2}d\Omega ^{2}\text{.}
\end{equation}%
To turn this form of the metric to an explicitly synchronous form, we need to make
additional time reparametrisation, 
\begin{equation}
d\hat{t}=\frac{dT}{\sqrt{g}}
\end{equation}%
so that the metrics becomes 
\begin{equation}
ds^{2}=-d\hat{t}^{2}+gdy^{2}+T^{2}(\hat{t})d\Omega ^{2}.
\end{equation}%
The metric coefficients depend only on time, so this is a cosmology-like
metrics. Indeed, taken by itself it can be considered as a metric of the
Kantowski-Sachs (KS) Universe. The obvious cosmological intuition makes it
natural to treat particles with constant $y$ and angular coordinates as
being at rest and consider peculiar velocities with respect to them.

The geodesic equations give us for the  4-velocity components
\begin{equation}
u^{\hat{t}}=\frac{d\hat{t}}{d\tau }=\frac{P}{\sqrt{g}}, \label{ttau}
\end{equation}%
\begin{equation}
u^{y}=\frac{dy}{d\tau }=-\frac{\varepsilon }{g},  \label{ye}
\end{equation}%
\begin{equation}
u^{\phi }=\frac{d\phi }{d\tau }=\frac{\mathcal{L}}{T^{2}}. \label{fi2tau}
\end{equation}%
As usual, 
\begin{equation}
u_{\mu }u^{\mu }=\frac{\mathcal{L}^{2}}{T^{2}}+\frac{\varepsilon ^{2}}{g}-%
\frac{P^{2}}{g}=-1.
\end{equation}%
We see from (\ref{ye}) that the coordinate $y$ is constant if $\varepsilon
=0 $. We are familiar with this condition from the Sec.VI where we
have shown 
(see the discussion after Eq.(64)) that such particles have peculiar velocities \textit{with respect
to the Lema\^{\i}tre frame} equal to the inverse flow velocity of
the Lema\^{\i}tre frame. In the Kantowski-Sachs frame considered in the
present section, the peculiar velocities of particles with $\varepsilon =0$
are zero by definition.

For the general situation, we fix the tetrad in the coordinates ($\hat{t},y,\theta
,\phi $) as
\begin{equation}
\hat{h}_{(0)}^{\mu }=(1,0,0,0),
\end{equation}%
\begin{equation}
\hat{h}_{(0)\mu }=-(1,0,0,0),
\end{equation}%
\begin{equation}
h_{(1)}^{\mu }=(0,\frac{1}{\sqrt{g}},0,0),
\end{equation}%
\begin{equation}
\hat{h}_{(1)\mu }=(0,\sqrt{g},0,0),
\end{equation}%
\begin{equation}
\hat{h}_{(3)\mu }=(0,0,0,\left\vert T\right\vert ).
\end{equation}

A simple calculation using (\ref{ttau}) - (\ref{fi2tau}) gives 
\begin{equation}
\hat{V}^{(1)}=\frac{\hat{h}_{(1)\mu }u^{\mu }}{-\hat{h}_{(0)\mu }u^{\mu }}=%
\sqrt{g} \frac{dy}{d\hat{t}}=-\frac{\varepsilon }{P},
\label{eZ}
\end{equation}%
\begin{equation}
\hat{V}^{(3)}=\frac{\mathcal{L}\sqrt{g}}{\left\vert T\right\vert P}.
\end{equation}%
For the Schwarzschild metric and pure radial motion, our eq. (97) agrees with eqs. (18), (A.9) of \cite{pl}.

The absolute value of the peculiar velocity 
\begin{equation}
\hat{V}^{2}=\hat{V}^{(1)2}+\hat{V}^{(3)2}
\end{equation}%
can be presented as 
\begin{equation}
\hat{V}^{2}=\frac{1}{P^{2}}(\varepsilon ^{2}+\frac{\mathcal{L}^{2}}{T^{2}}g)=%
\frac{P^{2}-g}{P^{2}}
\end{equation}%
or, directly in terms of the conserved quantities 
\begin{equation}
\hat{V}^{2}=\frac{\frac{\varepsilon ^{2}}{g}+\frac{\mathcal{L}^{2}}{T^{2}}}{%
1+\frac{\varepsilon ^{2}}{g}+\frac{\mathcal{L}^{2}}{T^{2}}}.  \label{ve}
\end{equation}

If $\varepsilon =0$, evidently $\hat{V}^{(1)}=0$, and we have a simple
formula for the angular velocity 
\begin{equation}
\hat{V}^{(3)}=\frac{\mathcal{L}}{\sqrt{\mathcal{L}^{2}+T^{2}}}.  \label{v3L}
\end{equation}

If $\mathcal{L}=0$, $\hat{V}^{(3)}=0$, so that the condition for zero peculiar
velocity in terms of conserved quantities is $\varepsilon =0$ and $\mathcal{L}=0$.

It is easy to see that for any nonzero $\varepsilon$ near the horizon, where $g\rightarrow 0$,%
\begin{equation}
|\hat{V}^{(1)}|\rightarrow 1\text{, }\hat{V}^{(3)}\rightarrow 0
\end{equation}

On the other hand, near the singularity $g\rightarrow \infty $,
and if $\mathcal{L}=0$, then $P$ diverges as $\sqrt{g}$,
and it follows from (\ref{eZ}) that
\begin{equation}
\hat{V}^{(1)}\approx -\frac{\varepsilon }{\sqrt{g}}\approx - \frac{\varepsilon 
}{v}\rightarrow 0\text{.}  \label{v1s}
\end{equation}%
If $\mathcal{L}\neq 0$, then the asymptotics of $P$ near a singularity is different,
\begin{equation}
    P \approx \sqrt{g} \left|\frac{\mathcal{L}}{T}\right|, 
\end{equation}
so that
\begin{equation}
    V^{(1)} \approx -\frac{\varepsilon}{\sqrt{g}} \left|\frac{T}{\mathcal{L}}\right|
    \to 0.\label{v1L}
\end{equation}
For  the angular component we have%
\begin{equation}
\hat{V}^{(3)}\rightarrow sign\mathcal{L=}\pm 1\text{.}  \label{v3s}
\end{equation}

The "cosmological" intuition matches with these results completely. Since
the radial "scale factor" $g$ diverges, the radial peculiar velocity tends
to zero, while angular peculiar velocity tends to the speed of light since
the angular "scale factor" $T^{2}$ tends to zero. 
We see 
that near the singularity  the limiting values  of the radial and angular components of peculiar
velocity with respect to the Kantowski-Sachs frame (\ref{v1L}), (\ref{v3s})
are the same as with respect to the Lema\^{\i}tre frame (\ref{v1l}), (\ref{v3l}) though the forms of
asymptotics are different.

\section{Redshifts inside a black hole}

The redshift measured by an observer freely falling inside a black hole depends
rather nontrivially on an angular motion of photon and that of an observer
himself. We start with reminding a formal derivation of the redshift (for a full treatment see \cite{20}), and then explain how
different limiting cases can be understood intuitively using the formalism of
peculiar velocities. 

A photon is characterized by the wave vector $k_{\mu}$. Let us consider the
metric (\ref{met}). 
It is assumed that $k_{t}=-\omega_{0}$, where $\omega_{0}$ has the meaning of
the frequency measured at infinity. And, $k_{\phi}=-\omega_0 l$ has the meaning of the
angular momentum. A photon is propagating from  infinity inwards. 
The normalization condition $k_{\mu}k^{\mu}=0$ gives us outside the event
horizon in coordinates $(t,r,\phi)$%
\begin{equation}
k^{\mu}=(\frac{\omega_{0}}{f}, -Q\text{,}\frac{\omega_0 l}{r^{2}})\text{, }k_{\mu
}=(-\omega_{0}\text{, }\frac{ -Q}{f}\text{, }\omega_0 l)\text{,}\label{kf}%
\end{equation}
where,%
\begin{equation}
Q=\omega_0 \sqrt{1-\frac{fl^{2}}{r^{2}}}\text{.}\label{kl}%
\end{equation}

Below we will be interested in the properties of particles inside the
horizon. The corresponding formulas can be obtained by the substitution
$f=-g<0$. 

Under the horizon, we have in the coordinates $(T$, $y$,  $\phi$)
\begin{equation}
k^{\mu}=(Q,-\frac{\omega_{0}}{g},\frac{\omega_0 l}{T^{2}})\text{,}\label{k}%
\end{equation}%
\begin{equation}
k_{\mu}=(-\frac{Q}{g},-\omega_{0},\omega_0 l)\text{,}%
\end{equation}
where now $k_{y}=\omega_{0}$ is conserved, $\omega_{0}>0$,%
\begin{equation}
Q=\omega_0 \sqrt{1+\frac{g}{T^{2}}l^{2}}\text{.}\label{z}%
\end{equation}

It follows from (\ref{kl}) - (\ref{z}) that 
\begin{equation}
\omega =\frac{PQ}{g}-\omega _{0}  \frac{\varepsilon }{g}-\omega_0 l\frac{%
\mathcal{L}}{r^{2}}\text{.}  \label{Zv}
\end{equation}%
This general formula leads to a number of different asymptotics near a singularity, depending on 
motion of an observer an the photon observed, resulting in infinite redshifts, infinite blueshifts or finite redshifts. They are summarized in \cite{20}. Here we interpret these
results from the viewpoint of peculiar velocities.

Using (\ref{evp}) --  (\ref{v3v}) we obtain from (\ref{Zv})%
\begin{equation}
\omega =\frac{
Q(v-V^{(1)})- \omega_{0}(1-vV^{(1)})}
{g\sqrt{1-v_{p}^{2}}}
-\frac{\omega_0 l}{r}\frac{V^{(3)}}{\sqrt{%
1-v_{p}^{2}}}.
\end{equation}

Let $l=0$. Then $Q=\omega _{0}$ and%
\begin{equation}
\frac{\omega }{\omega _{0}}=
\frac{(v-V^{(1)})- (1-vV^{(1)})}{g \sqrt{1-v_{p}^{2}}}.%
\end{equation}%
Substituting  (\ref{vf}), we obtain%
\begin{equation}
\frac{\omega }{\omega _{0}}=
\frac{(1+ V^{(1)})}{(1+ v)\sqrt{%
1-v_{p}^{2}}}.
\end{equation}%

If the motion is purely radial, $V^{1}=V$, $v_{p}^{2}=V^{2}$, we return to our
known formula 11.3 from \cite{jcap} (eq. 78 of its arxiv version)%
\begin{equation}
\frac{\omega }{\omega _{0}}=\frac{\sqrt{1+ V}}{(1+ v)\sqrt{%
1- V}}.
\end{equation}

{The peculiar velocity notation gives us the possibility to interpret some
results on redshifts observed by a freely falling observer near the
singularity. Indeed, the combined redshift is 
\begin{equation}
(1+z)=(1+z_{g})(1+z_{d}),  \label{1+z}
\end{equation}%
where } 
\begin{equation}
1+z_{g}=\frac{1}{1+v}\text{, }1+z_{d}=\frac{\sqrt{1+ V}}{%
\sqrt{1- V}}\text{.}
\end{equation}%
The gravitational part $z_{g}$ is the shift observed by a particle in the
flow (with a zero peculiar velocity), which has been calculated in the Lema%
\^{\i}tre frame in \cite{we}, and $z_{d}$ is the Doppler shift due to a nonzero
peculiar velocity. In this interpretation equation (\ref{1+z}) represents an
evident combination of the Lorentz redshift (the part which depends on $V$%
) and the gravitational redshift (the part with depends on $v$). Near a
singularity, depending on the direction of the photon, we have either
infinite redshift or infinite blueshift.

The same decomposition can be written also with respect to the KS frame (87), (92)-(96). We continue to consider a radial photon and a radially falling observer ($l=0$, $\mathcal{L}=0$). Now 
the redshift measured by an observer in the flow ($\varepsilon=0$) has even simpler form since
only the first term in the main formula (\ref{Zv}) is non-zero: $\omega/\omega_0=1/\sqrt{g}$.
For an observer with non-zero $\varepsilon$ we get 
\begin{equation}
    \frac{\omega}{\omega_0}=\frac{P- \varepsilon}{g}.
\end{equation}
Using (5) with $\mathcal{L}=0$ and (\ref{eZ}), it can be rewritten in the form
\begin{equation}
    \frac{\omega}{\omega_0} =\frac{1}{\sqrt{g}}\frac{\sqrt{1+ \hat{V}^{(1)}}}{\sqrt{1- \hat{V}^{(1)}}}
\end{equation}
giving again a combination of gravitational and Lorentz redshifts, but with respect to the KS
frame now.

Now we consider a situation with a nonradial photon ($l \ne 0$). It is easier to work in the KS frame.
For a radially falling observer  near the singularity  we have $g \to \infty$, and recalling
the definitions of $P$ and $Q$ we see that the first term in (\ref{Zv}) dominates giving an infinite blueshift
with the asymptotic
\begin{equation}
    \frac{\omega}{\omega_0} \approx  \frac{l}{|T|}.
    \label{lT}
\end{equation}
Let us now consider  
the case of a nonradial motion of an observer as well. In this case the infinite gravitational blueshift (\ref{lT})
is combined with an infinite Doppler redshift since the
angular peculiar velocity approaches 1. Indeed, near the singularity, the
term with $\varepsilon $ in (\ref{ve}) is negligible, $V^{(1)}\rightarrow 0$
according to (\ref{v1s}), so eq. (\ref{v3L}) becomes a good approximation.
It can be rewritten as%
\begin{equation}
\mathcal{L}=\frac{\left\vert T\right\vert \hat{V}^{(3)}}{\sqrt{%
(1-\hat{V}^{(3)})(1+\hat{V}^{(3)})}}\text{.}
\end{equation}

{Taking into account that now }$T\rightarrow 0$, {for a finite }$\mathcal{L}$
the velocity component $\left\vert \hat{V}^{(3)}\right\vert \rightarrow 1$ in
agreement with (\ref{v3s}). Let for definiteness $\mathcal{L}>0$. Then{, }$%
\hat{V}^{(3)}\rightarrow +1$, and{\ 
\begin{equation}
\sqrt{1-\hat{V}^{(3)}}\approx \left\vert T\right\vert /(\mathcal{L}\sqrt{2})\text{.%
}
\end{equation}%
}

Let  $l>0$. Taking into account the Doppler factor, we find that the ratio
\begin{equation}
\frac{\omega}{\omega_0}=
\frac{l}{\left\vert T\right\vert }\sqrt{\frac{1-\hat{V}^{(3)}}{1+\hat{V}^{(3)}}}\approx 
\frac{l}{\left\vert T\right\vert }\frac{\left\vert T\right\vert }{\mathcal{L}%
\sqrt{2}\sqrt{2}}=\frac{l}{2L}  \label{lT+}
\end{equation}%
is finite and coincides with the result obtained earlier in eq. (47) of \cite%
{20}.

If $l<0$, the velocities of an observer and a photon are pointed in opposite
directions. Then, instead of (\ref{lT+}) we have%
\begin{equation}
\frac{\omega}{\omega_0}=
\frac{\left\vert l\right\vert }{\left\vert T\right\vert }\sqrt{\frac{%
1+\hat{V}^{(3)}}{1-\hat{V}^{(3)}}}\approx \frac{2\mathcal{L}\left\vert l\right\vert }{%
T^{2}}\rightarrow \infty \text{.}
\end{equation}%
This agrees with eq. (48) of \cite{20}.

\ 

\section{Conclusions}
In the present paper we have considered the 3-velocity of an object falling freely into a
black hole with respect to two different freely falling frames. By direct analogy with the
cosmological terminology, we call this 3-velocity a peculiar velocity. We determined the
dependence of peculiar velocities components of freely falling objects on integrals of motion
and considered their asymptotics near horizons and a singularity. We have developed rather a general
approach that can, in principle, be applied both in the cosmology and physics of black holes,
including their interiors. Now, the concept of a peculiar velocity is exploited to include
nonradial motion. This, in particular, enabled us to give a simple qualitative explanation of the
phenomenon of  red/blue shifts inside a black hole, especially near the singularity. It agrees
with the results of direct calculations done earlier. We have also shown how the general formula for
the frequency shift in a radial fall admits a simple decomposition to the gravitational and
kinematic parts for two considered frames.
Since we considered geodesic motion in a fixed static spherically symmetric metric (without
demanding that this metric is a solution of General Relativity equations), our results are
valid for any black hole solutions of the form (1) in any metric theory with geodesic motion
of particles and photons.

\section*{Acknowledgements}
\bigskip The work was supported by the Russian Government Program of
Competitive Growth of Kazan Federal University.

\end{document}